\documentclass[aps,prd,superscriptaddress,nofootinbib,amsmath,amsfonts,preprintnumbers,notitlepage,10pt,english]{revtex4}
\usepackage{amsmath}
\usepackage{amssymb}
\usepackage{babel}
\usepackage{graphicx}
\usepackage{color}

\newcommand{\f}[2]{\frac{#1}{#2}}
\newcommand{\mk}[1]{\left( #1 \right)}
\newcommand{\kk}[1]{\left[ #1 \right]}

\newcommand{\be}{\begin{equation}}
\newcommand{\ee}{\end{equation}}
\newcommand{\bea}{\begin{eqnarray}}
\newcommand{\eea}{\end{eqnarray}}

\def\U{{\cal U}}
\def\K{{\cal K}}
\def\Mpl{M_{\rm Pl}}



\begin{document}

\preprint{RESCEU-41/12}

\title{Self-accelerating solutions in massive gravity on an isotropic reference metric}

\author{Hayato Motohashi}
\affiliation{Department of Physics, Graduate School of Science,
The University of Tokyo, Tokyo 113-0033, Japan}
\affiliation{Research Center for the Early Universe (RESCEU),
Graduate School of Science, The University of Tokyo, Tokyo 113-0033, Japan}

\author{Teruaki Suyama}
\affiliation{Research Center for the Early Universe (RESCEU),
Graduate School of Science, The University of Tokyo, Tokyo 113-0033, Japan}

\begin{abstract}
Within the framework of the recently proposed ghost-free massive gravity,  
a cosmological constant-type self-accelerating solution has been obtained for Minkowski
and de Sitter reference metrics.
We ease the assumption on the reference metric and find the self-accelerating solution
for the reference metric respecting only isotropy, 
thus considerably extending the range of known solutions.
\end{abstract}

\date{\today}

\maketitle

\section{Introduction}

It is interesting to consider the possibility that the graviton has a nonzero mass, from not only a theoretical, but also a phenomenological point of view. It is intuitively expected that in the presence of a nonzero graviton mass $m$, the gravitational potential has the Yukawa form $\sim e^{-mr}/r$. The gravitational force then decays at a scale larger than $m^{-1}$, and it could be the origin of the accelerated expansion of the Universe. Therefore, a theory of massive gravity may be an alternative to dark energy.

The history of the challenges to explore massive gravity begins from the first attempt made by Fierz and Pauli~\cite{Fierz:1939ix}. They introduced the graviton mass term in the action. 
Later, it was revealed that at the linear level, the Solar System-scale prediction in the massless limit of their theory is not in agreement with that in general relativity, which is called the 
van Dam--Veltman--Zakharov discontinuity~\cite{vanDam:1970vg,Zakharov:1970cc}. 
This discontinuity actually can be cured if we consider the nonlinear effect~\cite{Vainshtein:1972sx}. 
In the vicinity of a massive object, the nonlinear effect provides a larger contribution than the linear effect. 
This screening scale is called the Vainshtein radius. 
Inside the Vainshtein radius, the nonlinear effect is responsible for screening the additional degrees of freedom from general relativity.
At the same time, however, the nonlinearity brought another problem. There exist the ghost degrees of freedom, which show up at the nonlinear level~\cite{Boulware:1973my}.
Due to this Boulware-Deser ghost, it has been difficult to achieve a healthy theory of massive gravity.

Recently, de Rham, Gabadadze, and Tolley proposed how to construct the Lagrangian of massive gravity, which is free of the Boulware-Deser ghost in the decoupling limit on the Minkowski reference metric~\cite{deRham:2010ik,deRham:2010kj}.
This theory introduces another metric in addition to the physical metric, which is referred as to the reference or fiducial metric.
The absence of the Boulware-Deser ghost at the nonlinear level was proved \cite{Hassan:2011hr,Hassan:2011ea} even on the general reference metric~\cite{Hassan:2011tf,Hassan:2012qv}.
After the ghost-free massive gravity theory was established, 
cosmological solutions in this theory were constructed \cite{deRham:2010tw,D'Amico:2011jj,Koyama:2011xz,Koyama:2011yg,Gumrukcuoglu:2011ew,Volkov:2011an,Gratia:2012wt,Kobayashi:2012fz,Langlois:2012hk} for Minkowski or de Sitter reference metrics (See also Ref.~\cite{Fasiello:2012rw} for cosmological perturbations on the Friedmann-Lema\^itre-Robertson-Walker reference metric).

In this paper, we derive cosmological constant-type self-accelerating solutions in the above ghost-free massive gravity on the general isotropic reference metric characterized by two arbitrary functions [see Eq.~(\ref{reference-metric})]. 

\section{Massive gravity on the isotropic reference metric}
\label{sec-be}

The ghost-free massive gravity is defined by the Lagrangian with graviton mass $m$~\cite{deRham:2010ik,deRham:2010kj},
\be {\cal L}_{\rm MG}=\f{M^2_{\rm Pl}}{2}\sqrt{-g}\mk{R-\f{m^2}{4}\U}. \ee
The potential $\U$ is given by two parametric form 
\be -\f{\U}{4}={\cal L}_2+\alpha_3 {\cal L}_3+\alpha_4 {\cal L}_4, \ee
and each part is 
\bea
{\cal L}_2&=& [\K]^2-[\K^2],\\
{\cal L}_3&=& [\K]^3-3[\K][\K^2]+2[\K^3],\\
{\cal L}_4&=& [\K]^4-6[\K]^2[\K^2]+8[\K][\K^3]+3[\K^2]^2-6[\K^4],
\eea
where the brackets represent traces and $\K^\mu_\nu = \delta^\mu_\nu-\sqrt{\Sigma}^\mu_\nu$. 
$\Sigma^\mu_\nu$ is defined by 
\be \Sigma^\mu_\nu = g^{\mu\rho}\partial_\rho \phi^a \partial_\nu \phi^b f_{ab}, \ee
where $g_{\mu\nu}$ is the physical metric, $f_{ab}$ is the reference metric, and $\phi^a$ is the St\"uckelberg field.

We consider the following isotropic forms for the physical metric, the reference metric, and the St\"uckelberg field:
\bea
&&g_{\mu\nu}dx^\mu dx^\nu=-N^2(t,r)dt^2+a^2(t,r)\delta_{ij}dx^i dx^j,\\
&&f_{ab}\partial_\mu \phi^a \partial_\nu \phi^b=-n^2(\phi^0,\sqrt{\phi^i \phi^i})\partial_\mu \phi^0 \partial_\nu \phi^0+\alpha^2(\phi^0,\sqrt{\phi^i \phi^i})\delta_{ij}\partial_\mu \phi^i \partial_\nu \phi^j, \label{reference-metric}\\
&&\phi^0=f(t,r),\quad \phi^i=g(t,r)\f{x^i}{r}.
\eea
Let us emphasize here that we do not assume any special forms for $n$ and $\alpha$.
So far, cosmological constant-type solutions have also been constructed on some fixed reference metric, {\it i.e.}, Minkowski~\cite{deRham:2010tw,D'Amico:2011jj,Koyama:2011xz,Koyama:2011yg,Gumrukcuoglu:2011ew,Volkov:2011an,Gratia:2012wt,Kobayashi:2012fz} or de Sitter~\cite{Langlois:2012hk}.
We also leave the forms of the St\"uckelberg fields unspecified.
Note that we refer to our ansatz as ``isotropic'' for the following reasons:
First, we fix the St\"uckelberg fields as an isotropic form with respect to the spacetime coordinates.
Next, if we regard the St\"uckelberg fields as coordinates, $f$ and $g$ are time and radial components, respectively. 
We set the reference metric to be isotopic with respect to such St\"uckelberg field coordinates.
Furthermore, since $f$ and $g$ themselves depend on $t$ and $r$, $n$ and $\alpha$ are actually determined by $t$ and $r$.
Thus, the reference metric is also isotropic in the above sense.
In the following discussion, we will derive an exact solution with these settings, which are more general than the analysis in the literature.

After the similar procedure described in Ref.~\cite{Gratia:2012wt}, the potential is rewritten as
\be \f{\U}{4}=P_0\mk{\f{\alpha g}{ar}}+\sqrt{X}P_1\mk{\f{\alpha g}{ar}}+WP_2\mk{\f{\alpha g}{ar}}, \ee
where 
\bea
P_0(x)&=&-12-2x(x-6)-12\alpha_3(x-1)(x-2)-24\alpha_4(x-1)^2,\\
P_1(x)&=&-2(2x-3)+6\alpha_3(x-1)(x-3)+24\alpha_4(x-1)^2,\\
P_2(x)&=&-2+12\alpha_3(x-1)-24\alpha_4(x-1)^2.
\eea
and 
\bea
X&=&\bigg(\f{n\dot f}{N}+\mu\f{\alpha g'}{a}\bigg)^2-\bigg(\f{\alpha \dot g}{N}+\mu\f{nf'}{a}\bigg)^2,\\
W&=&\mu \f{n \alpha}{N a}(\dot f g'-\dot g f'),
\eea
with $\mu$ being the sign function, $\mu={\rm sgn}(\dot f g'-\dot g f')$.

The equations of motion for the St\"uckelberg fields are then derived from variating the potential,
\bea
&&\f{\partial}{\partial t}\kk{\f{a^3r^2n}{\sqrt{X}}\bigg(\f{n\dot f}{N}+\mu\f{\alpha g'}{a}\bigg)P_1+\mu a^2 r^2 n\alpha g' P_2}
-\f{\partial}{\partial r}\kk{\f{a^2r^2Nn}{\sqrt{X}}\bigg(\mu\f{\alpha \dot g}{N}+\f{n f'}{a}\bigg)P_1+\mu a^2 r^2 n\alpha \dot g P_2} \notag \\
&&~~~~~~~=Na^3r^2\left[(P'_0+\sqrt{X}P'_1+WP'_2)\f{g}{ar}\f{\partial \alpha}{\partial f}+\bigg(\f{\dot f^2}{N^2}-\f{f'^2}{a^2}\bigg)\f{n P_1}{\sqrt{X}}\f{\partial n}{\partial f}+\bigg( \f{g'^2}{a^2}-\f{\dot g^2}{N^2}\bigg)\f{\alpha P_1}{\sqrt{X}}\f{\partial \alpha}{\partial f}\right. \notag \\
&&~~~~~~~~~~~~~~~~~~~~~~\left.+\mk{\f{P_1}{\sqrt{X}}+P_2}(\dot f g'-\dot g f')\f{\mu}{Na}\f{\partial}{\partial f}(n\alpha)\right], \label{eomf} \\
&&-\f{\partial}{\partial t}\kk{\f{a^3r^2\alpha}{\sqrt{X}}\bigg(\f{\alpha\dot g}{N}+\mu\f{n f'}{a}\bigg)P_1+\mu a^2 r^2 n\alpha f' P_2}
+\f{\partial}{\partial r}\kk{\f{a^2r^2N\alpha}{\sqrt{X}}\bigg(\mu\f{n \dot f}{N}+\f{\alpha g'}{a}\bigg)P_1+\mu a^2 r^2 n\alpha \dot f P_2} \notag \\
&&~~~~~~~=Na^3r^2\left[(P'_0+\sqrt{X}P'_1+WP'_2)\f{1}{ar}\f{\partial (\alpha g)}{\partial g}+\bigg(\f{\dot f^2}{N^2}-\f{f'^2}{a^2}\bigg)\f{n P_1}{\sqrt{X}}\f{\partial n}{\partial g}+\bigg( \f{g'^2}{a^2}-\f{\dot g^2}{N^2}\bigg)\f{\alpha P_1}{\sqrt{X}}\f{\partial \alpha}{\partial g}\right. \notag \\
&&~~~~~~~~~~~~~~~~~~~~~~\left.+\mk{\f{P_1}{\sqrt{X}}+P_2}(\dot f g'-\dot g f')\f{\mu}{Na}\f{\partial}{\partial g}(n\alpha)\right]. \label{eomg}
\eea
We find that an exact solution for this system is given by setting $\alpha g/ar = x_0$, where $P_1(x_0)=0$, namely, 
\be x_0=\f{1+6\alpha_3+12\alpha_4\pm\sqrt{1+3\alpha_3+9\alpha_3^2-12\alpha_4}}{3(\alpha_3+4\alpha_4)}. \ee
This condition fixes $g$ in terms of $\alpha$ and $a$.
The concrete expression of $g$ depends on the explicit form of $\alpha$. 
Equations \eqref{eomf} and \eqref{eomg} then read
\bea
g'\f{\partial}{\partial t}(n\alpha a^2 r^2)-\dot g \f{\partial}{\partial r}(n\alpha a^2 r^2)
&=&\f{Na^2rg}{\mu P_2}\f{\partial \alpha}{\partial f}(P'_0+\sqrt{X}P'_1+WP'_2)+a^2r^2(\dot f g'-\dot g f')\f{\partial}{\partial f}(n\alpha),\\
-f'\f{\partial}{\partial t}(n\alpha a^2 r^2)+\dot f \f{\partial}{\partial r}(n\alpha a^2 r^2)
&=&\f{Na^2r}{\mu P_2}\f{\partial (\alpha g)}{\partial g}(P'_0+\sqrt{X}P'_1+WP'_2)+a^2r^2(\dot f g'-\dot g f')\f{\partial}{\partial g}(n\alpha).
\eea
We can derive the time and spatial derivative of $(n\alpha a^2 r^2)$ in the matrix form,
\be
\left(
\begin{array}{c}
\displaystyle \f{\partial}{\partial t}(n\alpha a^2 r^2) \\
\displaystyle \f{\partial}{\partial r}(n\alpha a^2 r^2)
\end{array}
\right)
=
\left(
\begin{array}{cc}
\dot f & \dot g \\
f' & g'
\end{array}
\right)
\left(
\begin{array}{c}
\displaystyle \f{Na^2rg}{\mu P_2(\dot f g' - \dot g f')}\f{\partial \alpha}{\partial f}(P'_0+\sqrt{X}P'_1+WP'_2)+a^2r^2\f{\partial}{\partial f}(n\alpha) \\
\displaystyle \f{Na^2r}{\mu P_2(\dot f g' - \dot g f')}\f{\partial (\alpha g)}{\partial g}(P'_0+\sqrt{X}P'_1+WP'_2)+a^2r^2\f{\partial}{\partial g}(n\alpha)
\end{array}
\right).\label{tandr}
\ee
The first line is 
\be \f{\partial}{\partial t}(n\alpha a^2 r^2)=\f{Na^2r}{\mu P_2(\dot f g' - \dot g f')}(P'_0+\sqrt{X}P'_1+WP'_2)\f{\partial}{\partial t}(\alpha g)+a^2r^2\f{\partial}{\partial t}(n\alpha). \ee
By substituting $g=x_0ar/\alpha$, we obtain
\be  \f{2\mu nP_2[\dot f (ar)' - f'(ar)\dot{}~]}{\displaystyle 1+\f{x_0ar}{\alpha^2}\f{\partial \alpha}{\partial g}}=Na(P'_0+\sqrt{X}P'_1+WP'_2). \label{eomff} \ee
The second line of Eq.~\eqref{tandr} also yields the same equation. 
Thus, the St\"uckelberg field $f$ is determined as a solution of Eq.~\eqref{eomff}.

For the above solution, the energy momentum tensor for the potential $\displaystyle T^{(\U)}_{\mu\nu}\equiv\f{\Mpl^2m^2}{\sqrt{-g}}\f{\delta}{\delta g^{\mu\nu}}\f{\sqrt{-g}~\U}{4}$ takes a cosmological constant form:
\be
T^{(\U)\mu}_{~~~~~\nu}=
\left(
\begin{array}{cccc}
-\rho_\U & 0 & 0 & 0 \\
0 & P_\U & 0 & 0 \\
0 & 0 & P_\U & 0 \\
0 & 0 & 0 & P_\U
\end{array}
\right), \quad \rho_\U=-P_\U=\f{\Mpl^2m^2}{2}P_0(x_0).
\ee
Therefore, massive gravity on the general isotropic reference metric always has
two cosmological constant-type self-accelerating solutions.

Similarly to the case of Ref.~\cite{Gratia:2012wt}, our metric ansatz allows for the existence of isotropic distribution of matter. The Universe undergoes matter-dominated regime, which is smoothly followed by a regime dominated by the effective fluid with a cosmological constant form. Therefore, our solution can describe the same expansion history as that in the ${\rm \Lambda CDM}$ model.

\section{Conclusion}
\label{sec-cn}

We have derived the exact solution in ghost-free massive gravity on the isotropic reference metric
given by Eq.~(\ref{reference-metric}) where $n$ and $\alpha$ are arbitrary functions. 
The solution has an energy momentum tensor of cosmological constant type. The derivation of the solution does not rely on the ansatz for the physical and reference metric or the St\"uckelberg field, apart from their isotropy. Therefore, the massive gravity-induced fluid, which behaves like a cosmological constant, can coexist with the isotropically distributed matter. It could be an alternative to dark energy.

\begin{acknowledgments}
This work was supported by JSPS Research Fellowships for Young Scientists (H.M.) and 
Grant-in-Aid for JSPS Fellows No.~1008477 (T.S.).
\end{acknowledgments}

\end{document}